\documentclass[12pt, twoside]{article}
\usepackage{graphicx}
\usepackage{hyperref}

\begin{document}

\begin{center}
{\Large\bf Coordinate Geometric Approach to Spherometer}\footnote{\normalsize 
Published in the  
{\em Bulletin of the IAPT}, {\bf 5}(6), pp. 139-142 (June 2013). 
({\bf IAPT}: Indian Association of Physics Teachers). 
\url{http://indapt.org/} and \url{http://www.iapt.org.in/}. 
} \\

~\\

\noindent
Sameen Ahmed Khan \\
Engineering Department \\
Salalah College of Technology ({\bf SCOT}) \\
Post Box No. 608, Postal Code: 211 \\
Salalah, {\bf Sultanate of Oman}. \\
rohelakhan@yahoo.com, 
\url{http://SameenAhmedKhan.webs.com/}
\end{center}

%\medskip

\begin{abstract}
The spherometer used for measuring radius of curvature of spherical surfaces is 
explicitly based on a geometric relation unique to circles and spheres.  
We present an alternate approach using coordinate geometry, which reproduces the 
well-known result for the spherometer and also leads to a scheme to study aspherical surfaces.  
We shall also briefly describe some of the modified spherometers. 
\end{abstract}




%\newpage

\section{Introduction}
The focal length of spherical lenses is governed by the Lens-Maker's Equation, 
which contains the radii of both surfaces of the lens~\cite{Anderson-Burge}.  
Spherometers are precision instruments, which were designed by the opticians of 
the early-nineteenth century (or even earlier) to deduce the radius of curvature 
of spherical surfaces, as the name suggests~\cite{Bud-Warner}.  
The initial design of a three-legged base supporting a central micrometer screw, 
was quickly adopted as the basic standard and remains in use to this 
day~\cite{Introductory-1}-\cite{Khan-lab}.  
\begin{figure}[ht]
\begin{center}
\includegraphics[width=8cm]{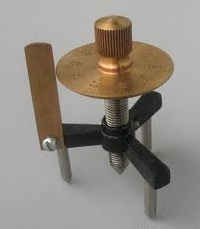} 
\caption{The Common Spherometer.}
\label{figure-common-spherometer}
\end{center}
\end{figure}
The error arising from the deviations 
from the tripod design have been studied~\cite{Trikha-Bhatia}.  
The need to handle aspherical surfaces (non-spherical surfaces) has resulted in 
modifications of the traditional spherometers~\cite{Rowell}.  One of the most 
prominent modifications resulted in the invention of the cylindrometer, a dual 
device, which can additionally measure the radius of curvature of a right circular 
cylinder~\cite{Cylindrometer-Singh}-\cite{Khan-cylindrometer-iapt}.  
The working of both the spherometer and the cylindrometer (also known as the {\em Cylindro-Spherometer} 
and {\em Sphero-Cylindrometer}) are based on a geometric relation unique to circles and spheres, 
dating back to the time of Euclid.  In this article, we present an alternate derivation of the 
spherometer formula using coordinate geometry, which reproduces the familiar result for the 
spherometer.  This approach, using the powerful techniques of coordinate geometry is suitable to 
a generalization of the traditional spherometer to devices, which can characterize aspherical 
surfaces~\cite{Khan-quadratic-surfaces}.  We shall first review the traditional derivation of 
the spherometer formula and then describe the coordinate geometric approach.  
An Appendix is dedicated to the variants and modifications of the common spherometer.

\section{Spherometer: the Traditional Approach} 
\begin{figure}[ht]
\begin{center}
\includegraphics[width=8cm]{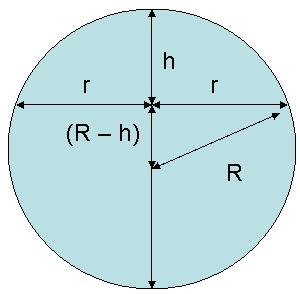} 
\caption{The Geometry of the Spherometer.}
\label{figure-spherometer-euclidean-geometry}
\end{center}
\end{figure}

The commonly available spherometers consist of a tripod framework supported on 
three fixed legs of equal lengths.  The tips of the three legs lie on the corners 
of an equilateral triangle of side $L$.  An accurately cut micrometer-screw passes 
through the nut fixed at the centroid of the equilateral triangle.  
The micrometer-screw is parallel to the three fixed legs.  
A large circular disc with typically a hundred divisions is attached to the top 
of the micrometer-screw.  A small millimeter scale is vertically attached to the 
tripod framework (parallel to the axis of the micrometer-screw).  
The two scales together (working on the principle of the screw gauge) 
provide the relative height of the micrometer-screw (known as {\em sagitta}) with 
respect to the tips of the fixed legs.  When the spherometer is placed on a 
spherical surface, the underlying geometry is a circle with the equilateral triangle 
inscribed in it.  The great circle (whose radius is $R$) touches the tip of 
the micrometer-screw of height $h$, measured relative to the plane containing the 
tips of the three fixed legs, as seen 
in Fig.-\ref{figure-spherometer-euclidean-geometry}.  
Applying the Pythagorean theorem in Fig.-\ref{figure-spherometer-euclidean-geometry}, 
we have $R^2 = r^2 + (R - h)^2$, leading to 
\begin{eqnarray}
R(h) & = & \frac{r^2}{2 h} + \frac{h}{2}\,, \nonumber \\
h(r) & = & R - \sqrt{R^2 - r^2} = \frac{r^2}{R + \sqrt{R^2 - r^2}}\,.
\label{spherometer-basic}
\end{eqnarray}
The radius, $r$ of the circumcircle is related to the side, $L$ of the equilateral triangle 
by the relation $r = L/\sqrt{3}$.  So, the radius of the sphere is given by 
\begin{eqnarray}
R(h) = \frac{L^2}{6 h} + \frac{h}{2}\,.
\label{spherometer-sphere}
\end{eqnarray}

\section{Coordinate Geometric Approach to \\ the Spherometer}
A sphere is uniquely determined by specifying four points on 
it~\cite{Shanti-Narayan}-\cite{Weisstein-Mathworld}.  This is precisely the situation 
in a spherometer: the three points are the fixed tips of the tripod and the fourth 
point is the movable tip of the micrometer-screw.  It is possible to have a 
coordinate representation of the spherometer as follows.  The tips of the three 
fixed legs (lying on the corners of the equilateral triangle of side $L$) and the tip 
of the movable micrometer-screw (lying at the centroid of the equilateral triangle), 
on a plane surface can be chosen to lie completely in the $X$-$Y$ plane, without any 
loss of generality.  The tip of the micrometer-screw moves parallel to the $Z$-axis.  
\begin{figure}[ht]
\begin{center}
\includegraphics[width=8cm]{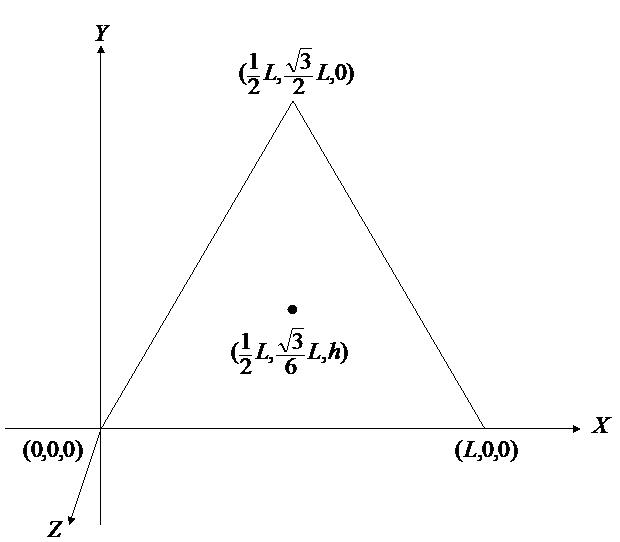} 
\caption{A Coordinate Representation of the Spherometer.}
\label{figure-spherometer-coordinate-geometry}
\end{center}
\end{figure}
In this choice of representation, when the spherometer is placed on a plane surface, 
the $z$-coordinates of all the four points are identically zero.  We call this as 
the {\em ground state} of the spherometer.  When the spherometer is placed on a spherical 
surface, only the $z$-coordinate of the micrometer-screw changes to $h$ and the 
remaining three points remain unchanged.  
 
The general equation of a sphere has four independent constants and is 
\begin{eqnarray}
x^2 + y^2 + z^2 + 2 u x + 2 v y + 2 w z + d = 0\,. 
\label{sphere-equation}
\end{eqnarray}
The radius of this sphere is $R = \sqrt{u^2 + v^2 + w^2 - d}$ and its 
centre is at $(- u, - v, - w )$.  
The four constants in the above equation are 
determined uniquely from the four points lying on the sphere.  It is to be noted 
that the four points can not be coplanar.  Using the three points of the tips of the 
fixed legs (inbuilt in the construction of the spherometer and the choice of the 
coordinate system), we obtain $d = 0$, $u = (- 1/2)L$, and $v = (-\sqrt{3}/6)L$.  
The coordinates of the micrometer-screw lead to $w = (L^2/6h) - h/2$.  
Substituting the values of these four constants, the radius of the sphere is readily 
obtained 
\begin{eqnarray}
R = \sqrt{u^2 + v^2 + w^2 - d} = \frac{L^2}{6 h} + \frac{h}{2}\,.
\label{spherometer-sphere-cg}
\end{eqnarray}
Thus, we reproduce the familiar result using coordinate geometry.  The choice of the 
coordinate representation ensures that the equation of the sphere is dependent only 
on one measurable parameter, namely the height $h$ of the screw.  The central feature 
of the coordinate geometric approach is that it does not use the underlying geometric 
relation explicitly, as is customary in the traditional approach.  Moreover, it is 
suited to incorporate the changes in the geometry of the spherometer.  For instance, 
during construction or during operation (particularly, when the spherometer is large and heavy) 
the device may deviate from the equilateral triangle geometry.  In which a case one only 
need to revise the ground state of the spherometer (which can be done even during 
the measurements) and then solve for the equation of the sphere.  Another advantage of 
using the coordinate geometric approach is that it offers a scheme to generalize the 
traditional spherometer to devices for studying aspherical surfaces.

\section{Concluding Remarks}
In the language of coordinate geometry, a sphere is uniquely determined by specifying 
four points on it.  In the spherometer the four points are the tips of the tripod and 
the movable tip of the micrometer-screw.  A right circular cylinder is uniquely determined 
by specifying five points on it; consequently the cylindrometer is a five point device.  
The tips of four fixed legs on a square base and the movable tip of the micrometer-screw 
at the centre account for the five points.  
Other surfaces such as cylindrical lenses or paraboloidal reflectors arise naturally 
in optics.  The study of elliptic and hyperbolic mirrors dates back to the time of 
Greeks~\cite{Ghatak, Hecht} and the medieval Arabs~\cite{Khan-eps, Khan-opn}.  
This has necessitated the study of a general class of surfaces known as the quadratic surfaces.  
In the context of optics the quadratic surfaces have been classified by enumerating the 
Hamiltonian orbits~\cite{Simon-1}-\cite{Wolf}. 
A quadratic surface is described by the general second-order equation, which has nine 
independent constants~\cite{Shanti-Narayan}-\cite{Weisstein-Mathworld}.  
The nine points (no four points being coplanar) uniquely determine the quadratic surface. 
So, we require a nine point device to identify any quadratic surface.  The four-point 
coordinate representation of a spherometer, has been generalized to a nine-point device, 
the quadricmeter, which generates the equation of the quadratic surfaces in terms of 
measurable laboratory parameters~\cite{quadricmeter}.  The quadricmeter is the instrument 
devised to identify, distinguish and measure the various characteristics (axis, foci, 
latera recta, directrix, {\em etc.},) completely characterizing the quadratic surfaces.  
The characterization is done using the standard techniques of coordinate geometry.  
One may use MS EXCEL~\cite{EXCEL}-\cite{EXCEL-IAPT} or a versatile symbolic package 
such as the MATHEMATICA incorporating the graphic environment~\cite{MATHEMATICA, MATHEMATICA-Boccara}.
In both the spherometer and the cylindrometer, one assumes the surface to be either 
spherical or cylindrical respectively.  In the case of the quadricmeter, there are 
no such assumptions.  The name {\em quadricmeter} originates from the word {\em quadrics} used for 
quadratic surfaces and was preferred over {\em conicoidmeter}. 

\setcounter{section}{0}
\section*{}
\renewcommand{\theequation}{A.{\arabic{equation}}}
\setcounter{equation}{0}

\begin{center}
{\large\bf Appendix: \\ 
Modifications of the Common Spherometer} \\
\end{center}

\noindent
Some of the widely used modifications of the common spherometer are: \\ %are described below. \\

\noindent
{\bf Ball Spherometer:} \\
Accuracy of measurement requires the legs of the spherometer to have sharp tips, 
which can damage the surfaces under study.  Blunt tips may save the surface but 
compromise on the accuracy.  This situation has been overcome in the ball spherometer.  
The tips of the legs are replaced by ball-bearings of radius $r_0$, modifying the basic relation to  
\begin{eqnarray}
R(h) = \frac{r^2}{2 h} + \frac{h}{2} \pm r_0\,,
\label{ball-spherometer}
\end{eqnarray}
where the positive and negative signs are for the convex and concave surfaces respectively. 

\bigskip

\noindent
{\bf Ring-Spherometer:} \\  
One of the widely used variations of the tripod spherometer is the ring-spherometer, 
which has a continuous ring of radius $r$ all the way around.  The radius, $R$ of the 
spherical surface is 
\begin{eqnarray}
R(h) = \frac{r^2}{2 h} + \frac{h}{2}\,.
\label{ring-spherometer}
\end{eqnarray}
Ring-spherometers measure up to the edge of a surface providing an average curvature.  
The ring needs to be fairly sharp at the edge or the ring will measure differently for 
concave and convex surfaces.  This can be corrected by using the internal and external 
radii of the ring for convex and concave surfaces respectively. 
\begin{figure}[ht]
\begin{center}
\includegraphics[width=8cm]{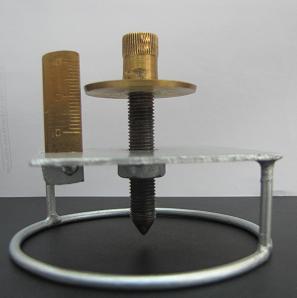} 
\caption{The Ring-Spherometer fabricated by the author.}
\label{figure-ring-spherometer}
\end{center}
\end{figure}

\bigskip

\noindent
{\bf Cylindrometer:} \\ 
The spherometer can be easily modified into a cylindrometer (also known as 
the {\em Cylindro-Spherometer} and {\em Sphero-Cylindrometer}) 
by replacing the tripod of the spherometer with a square framework supported on 
four fixed legs of equal lengths.  The tips of the four legs lie on the corners of a square 
of side $L$.  
\begin{figure}[ht]
\begin{center}
\includegraphics[width=8cm]{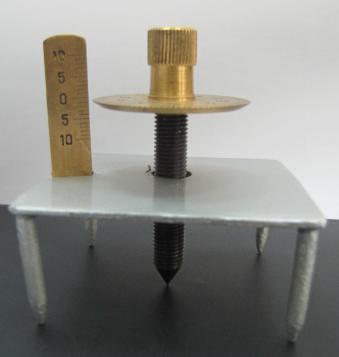} 
\caption{The Cylindrometer fabricated by the author.}
\label{figure-cylindrometer}
\end{center}
\end{figure}
The cylindrometer enables the measurement of the radius of a right circular cylinder in 
addition to the radius of spherical surfaces~\cite{Cylindrometer-Singh}-\cite{Khan-cylindrometer-iapt}.  
The corresponding expressions are 
\begin{eqnarray}
R_{\rm cylinder} & = & \frac{L^2}{8 h} + \frac{h}{2}\,,
\label{cylindrometer-cylinder} \\
R_{\rm sphere} & = & \frac{L^2}{4 h} + \frac{h}{2}\,.
\label{cylindrometer-sphere}
\end{eqnarray}

\bigskip

\noindent
{\bf Aspherical Surfaces:} \\
To some extent the aspherical surfaces can be analyzed using the common spherometer.  
In many situations, the aspheric surfaces can be approximated as conic sections of revolution.  
Incorporating the eccentricity, $e$ of the of the conic, Eq.~(\ref{spherometer-basic}) 
modifies to 
\begin{eqnarray}
h (r) = \frac{r^2}{R + \sqrt{R^2 - (1 - e^2) r^2}}\,.
\label{spherometer-aspheric}
\end{eqnarray}
Different conics have different eccentricities (for sphere $e = 0$; for prolate 
ellipsoid $0 < e < 1$, for oblate ellipsoid $e < 0$, for paraboloid, $e = 1$, and 
for hyperboloid $e > 1$).  A series expansion of Eq.~(\ref{spherometer-aspheric}) 
is used to calculate the aspheric departure~\cite{Anderson-Burge}.  A pair of readings 
obtained from two ring-spherometers of different sizes, close to the vertex can also 
be used to characterize such surfaces.

\end{document}